\documentclass[aps,prl,superscriptaddress,notitlepage,amsmath,amssymb,nofootinbib,reprint,longbibliography,preprintnumbers]{revtex4-1}
\usepackage[utf8]{inputenc} 
\usepackage{graphicx}
\usepackage{url}
\usepackage[bookmarks, pagebackref=false]{hyperref}
\usepackage[usenames,dvipsnames]{xcolor}
\usepackage[normalem]{ulem}

\newcommand{\ep}{\epsilon}
\newcommand{\as}{a_s}

\newcommand{\nf}{\ensuremath{n_f}}

\newcommand{\CF}{\ensuremath{C_F}}
\newcommand{\CA}{\ensuremath{C_A}}

\newcommand{\TF}{\ensuremath{T_F}}

\newcommand{\be}{\begin{equation}}
\newcommand{\ee}{\end{equation}}

\definecolor{bla}{HTML}{03396C}
\definecolor{blaa}{HTML}{005B96}
\definecolor{blaaa}{HTML}{6497B1}

\hypersetup{
  colorlinks,
  bookmarksopen,
  bookmarksnumbered,
  citecolor=blaa, 	
  linkcolor=blaa,    	
  urlcolor=blaa,		
}

\usepackage{natbib}

\allowdisplaybreaks

\begin{document}

\preprint{TTP24-033~P3H-24-061~TUM-HEP-1526/24~MPP-2024-179~ZU-TH 45/24}

\title{
  \Large\color{bla}
  Zero-jettiness soft function to third order in perturbative QCD}

\author{Daniel  Baranowski}\email{daniel.baranowski@physik.uzh.ch}
\affiliation{Physik Institut, Universität Zürich,
  Winterthurerstrasse 190, 8057 Zürich, Switzerland}
\author{Maximilian Delto}\email{maximilian.delto@tum.de}
\affiliation{Physics Department, Technical University of Munich,
  James-Franck-Strasse 1,  85748, Munich, Germany}
\author{Kirill  Melnikov}\email{kirill.melnikov@kit.edu}
\affiliation{Institute for Theoretical Particle Physics (TTP),
  Karlsruhe Institute of Technology, 76128, Karlsruhe, Germany}
\author{Andrey Pikelner}\email{andrey.pikelner@kit.edu}
\affiliation{Institute for Theoretical Particle Physics (TTP),
  Karlsruhe Institute of Technology, 76128, Karlsruhe, Germany}
\author{Chen-Yu Wang}\email{cywang@mpp.mpg.de}
\affiliation{Max-Planck Institute for Physics,
  Boltzmannstr.\ 8, 85748 Garching, Germany}

\begin{abstract}
We present the high-precision result for the zero-jettiness soft function at
next-to-next-to-next-to-leading order (N3LO) in perturbative QCD. At this
perturbative order, the soft function is the last missing ingredient required
for the computation of a hadronic colour singlet production or a colour singlet
decay into two jets using the zero-jettiness variable as the slicing parameter.
Furthermore, the knowledge of the N3LO soft function enables the re-summed
description of the thrust distribution in the process $e^+ e^- \to {\rm
hadrons}$ through next-to-next-to-next-to-leading logarithmic order, which is
important for the extraction of the strong coupling constant using this shape
variable. On the methodological side, the complexity of the zero-jettiness
variable forced us to develop a new semi-analytic method for phase-space
integration in the presence of constraints parameterized through Heaviside
functions which, hopefully, will be useful for further development of the
$N$-jettiness slicing scheme.
\end{abstract}
\maketitle

\section{Introduction}
\label{sec:intro}

During the next decade, experiments at the Large Hadron Collider (LHC) will test
the Standard Model (SM) of particle physics at an unprecedented level of
precision provided that uncertainties of the theoretical predictions can be
controlled. This opportunity resulted in an enormous effort in theoretical
collider physics, see Ref.~\cite{Heinrich:2020ybq} for a recent overview. Among
other things, these advances enabled fully-differential description of the Higgs
boson and the vector boson production processes at N3LO in perturbative
QCD~\cite{Chen:2022cgv,Neumann:2022lft,Campbell:2022uzw,Chen:2022lwc}.

Perturbative computations in QCD have to go beyond pure virtual amplitudes
because the latter are infra-red divergent. For infra-red safe observables,
these divergences are known to cancel with the real-emission contributions and,
in case of hadron collisions, with additional contributions that originate from
the collinear renormalization of parton distribution functions
~\cite{Bloch:1937pw, Kinoshita:1962ur,Lee:1964is}. Organizing such a
cancellation in an observable- and process-independent way is non-trivial. A
popular approach that allows one to do this is the so-called slicing method
where an observable $\delta$ is introduced to distinguish sub-processes with
Born kinematics ($\delta =0 $) from sub-processes with higher final-state
multiplicity ($\delta \ne 0$). The $\delta = 0$ case includes virtual
corrections, as well as unresolved soft and/or collinear real-emission
contributions. The universal factorization of the real-emission matrix elements
in the soft and collinear limits and the infra-red safety of the slicing
variable $\delta$ allows a simplified calculation of the $\delta=0$
contributions.

In practice, one introduces a cut-off parameter $\delta_c \ll 1$ and treats all
contributions with $\delta < \delta_c$ as unresolved, i.e. corresponding to the
$\delta=0$ case discussed above. Contributions from the phase-space regions with
$\delta > \delta_c$ require at least one formally resolved emission, so that the
perturbative order of the required computation drops by one unit compared to the
target precision of the inclusive cross section (i.e. N3LO becomes NNLO etc.).

Among several slicing variables that have been discussed in recent years, there
is just one that can be used to describe processes with arbitrary number of jets
beyond NLO. Dubbed $N$-jettiness in the original papers
\cite{Stewart:2009yx,Stewart:2010tn}, it was used as a slicing variable in
Refs.~\cite{Boughezal:2015dva,Gaunt:2015pea,Boughezal:2015aha,
Boughezal:2016wmq,Neumann:2022lft,Campbell:2023lcy} to compute NNLO QCD
corrections to several processes with either zero or one jet. For the
$N$-jettiness variables, the unresolved contribution can be expressed in terms
of the beam, soft and jet functions
\cite{Stewart:2009yx,Stewart:2010tn,Jouttenus:2011sg} which can be computed in
perturbative QCD independently of each other. The NNLO beam and jet functions
were computed in Refs.~\cite{Gaunt:2014cfa,Gaunt:2014xga, Boughezal:2017tdd}.
The NNLO soft functions for zero and one-jettiness were calculated in
Refs.~\cite{Kelley:2011ng,Monni:2011gb,
Boughezal:2015eha,Li:2016tvb,Campbell:2017hsw,Banerjee:2018ozf,Baranowski:2020xlp},
and for two-jettiness in Ref.~\cite{Jin:2019dho}. Extension of this approach to
higher-multiplicity processes requires computation of NNLO $N$-jettiness soft
function for \emph{arbitrary} number of hard final-state particles. This was
accomplished recently in Refs.~\cite{Bell:2023yso,Agarwal:2024gws}, making the
$N$-jettiness slicing scheme the \emph{very first perturbative framework} for
which, through NNLO, all unresolved real-emission ingredients are known
analytically for \emph{arbitrary} collider processes.

To extend the $N$-jettiness slicing scheme to N3LO in QCD we need to compute
beam, jet and soft functions through this perturbative order. While beam and jet
functions do not depend on the multiplicity $N$ and have been calculated a few
years ago
\cite{Bruser:2018rad,Ebert:2020lxs,Ebert:2020unb,Behring:2019quf,Baranowski:2022vcn},
the N$3$LO QCD soft functions remain unknown.

In fact, the computation of the N$3$LO $N$-jettiness soft function for an
arbitrary number of hard final-state partons is a problem of outstanding
technical complexity. Hence, as the first step it is reasonable to focus on the
simplest possible soft function, namely the one where the number of hard
emitters is exactly two. In this case, the $N$-jettiness variable reduces to the
zero-jettiness one, defined as follows
\begin{equation}
  \label{eq:Tau0def}
  \mathcal{T}_0(n) = \sum \limits_{i=1}^{n} \; {\text{min}} \left [
    \frac{2 p_a\cdot k_i}{P},\frac{2p_b \cdot k_i}{P}
  \right] \,.
\end{equation}
In Eq.~(\ref{eq:Tau0def}), the summation index runs over all \emph{soft} partons in a
particular process, $p_{a,b}$ are four-momenta of two hard emitters and $P$ is
an arbitrary variable with the mass dimension one.

The soft function is then computed with the help of the following formula
\begin{equation}
  S_\tau = \sum \limits_{n=0}^{\infty} S_\tau^{(n)},
\end{equation}
where
\begin{equation}
  \label{eq:StauEik}
  S_\tau^{(n)}
  = \frac{1}{{\cal N} }
  \int \prod \limits_{i=1}^{n} 
  \; 
  [\textrm{d} k_i] \; 
  \delta(\tau - {\cal T}_0(n) )
  \textrm{Eik} \left ( p_a,p_b,\{k\}_n
  \right ).
\end{equation}
In Eq.~(\ref{eq:StauEik}), $[{\rm d} k] = {\rm d}^{d}k/(2 \pi)^{d-1} \delta(k^2)
\theta(k^0)$ with $d = 4-2 \ep$, the eikonal function is defined as follows
\begin{equation}
  \label{eq:EikDef}
 {\rm Eik}  ( p_a,p_b,\{k\} )=    \lim_{\{k\} \to 0}^{}
    \frac{|{\cal M}(p_a,p_b,\{k\})|^2}{|{\cal M}(p_a,p_b)|^2},
\end{equation}
and ${\cal N}$ is the symmetry factor that is required  to properly account for the 
contribution of identical particles 
to $S^{(n)}_\tau$.

Each function $S_\tau^{(n)}$ describes a partonic process with $n$ final-state
soft partons computed to an arbitrary loop order, where also all loop amplitudes
have to be calculated in the leading soft approximation. Thus $S_\tau^{(n)} \sim
\alpha_s^n$ at the leading order and, thanks to the virtual corrections, it also
contains arbitrary powers of $\alpha_s$ in higher orders. A simple way to
visualize what needs to be computed is to recognize that $S_\tau$ is a
(normalized) cross section computed in the soft approximation for both real and
virtual partons, where the standard energy-momentum conservation condition is
replaced with the requirement that the zero-jettiness has a particular value
$\tau$.

\begin{figure*}[t]
  \centering
  \begin{align*}
    \includegraphics{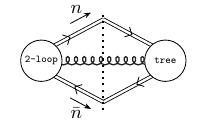}
    \includegraphics{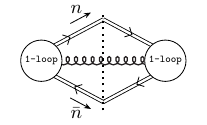}
    \includegraphics{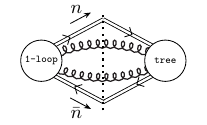}
    \includegraphics{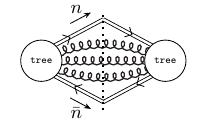}
  \end{align*}
  \caption{Different contributions to the zero-jettiness soft function at N3LO, see text for details. Only contributions with final-state gluons are shown.   Diagrams to the right of the cut are complex-conjugated. }
  \label{fig:diagsSF}
\end{figure*}

Hence, similar to the contributions to perturbative cross sections at high
orders of perturbation theory, the computation of the zero-jettiness soft
function at N$3$LO requires the triple-real contribution, the one-loop
correction to the double-real contribution and the two-loop correction to the
single-real contribution. We note that the triple-real contribution is known
partially \cite{Baranowski:2021gxe, Baranowski:2022khd}, whereas full results
are available for the other two \cite{Chen:2020dpk,Baranowski:2024ene}. While
the calculation of the one-loop correction to the double-real emission treat
phase-space and loop integration on the same footing, the simple form of the
\textit{integrated} two-loop correction to the single-emission soft
current~\cite{Duhr:2013msa} allows one to integrate over the soft-gluon phase
space \textit{after} the loop integration has been performed. Furthermore, the
three-loop correction to the Born amplitude vanishes in the soft approximation
since the appearing loop integrals are scaleless.

In this letter, we report on the completion of the calculation of the
triple-real contribution and present a high-precision result for the
zero-jettiness soft function to third order in perturbative QCD. In what
follows, we briefly describe the calculation summarizing technical challenges
and tests of the calculation that have been performed; a more detailed
discussion will be given elsewhere \cite{longpaper}. We present the result for
the N3LO contribution to the zero-jettiness soft function, which can also be
found in digital form alongside this submission, and conclude after that.

\section{Overview of the calculation}
\label{sec:calc}

The challenge with computing the soft function arises from the very definition
of the zero-jettiness variable, which, for each soft parton, requires selecting
the minimal value among scalar products of parton's and hard emitters' momenta.
Thus, this function can change abruptly in the middle of the phase space making
analytic integration of the soft function rather difficult. If, instead, one
attempts a numerical integration, one has to face a plethora of poorly
understood soft and collinear singularities, typical to a generic N3LO
computation, \emph{and} ultraviolet divergencies that arise because the
energy-momentum conservation requirement has been dropped. These problems make
the numerical calculation of the soft function highly non-trivial, especially in
such high an order as N3LO.

To enable the computation of the zero-jettiness soft function, we make use of
the fact that it is Lorentz invariant and its dependence on $P$, $s_{ab} = 2 p_a
\cdot p_b$ and $\tau$ can be easily reconstructed. Thus, we can set $\tau = 1$
and choose a reference frame and a factor $P$ without loss of generality. We
take $P = \sqrt{s_{ab}}$, and consider a center-of-mass frame where
\begin{equation}
p_{a} = \frac{\sqrt{s_{ab}}}{2} n,
\;\;\; p_b = \frac{\sqrt{s_{ab}}}{2} \bar n,
\end{equation}
with $n^2 = {\bar n}^2 = 0$ and
$n \cdot \bar n = 2$. We then write
\begin{equation}
k_i= \frac{\alpha_i}{2} n
+ \frac{\beta_i}{2} \bar n +
k_{\perp, i}.
\end{equation}
We use this decomposition and re-write Eq.~(\ref{eq:Tau0def}) as
\begin{equation}
{\cal T}_0(n) = \sum \limits_{i=1}^{n} {\text{min}} \left [
\alpha_i, \beta_i
\right],
\end{equation}
which allows us to project the phase-space integrations on the unique value of
${\cal T}_0$ by inserting partitions of unity
\begin{equation}
1 = \theta(\alpha_i - \beta_i)
+ \theta(\beta_i - \alpha_i),
\end{equation}
to the integrand in Eq.~(\ref{eq:StauEik}) for each of the soft partons.

Thus, to compute the triple-real contribution, we should integrate the eikonal
function in Eq.~(\ref{eq:EikDef}) over the phase space of three real gluons with a
$\delta$-function $\delta(\tau - x_1 -x_2 -x_3 )$ where $x_i = {\rm
min}[\alpha_i,\beta_i]$, and a product of three $\theta$-functions whose
arguments are $\pm (\alpha_i - \beta_i)$.

We note that the required eikonal function for three soft gluons was computed
and simplified in Ref.~\cite{Catani:2019nqv}. For the $q \bar q g$ soft
final-state partons the required eikonal function can be extracted from
Ref.~\cite{DelDuca:2022noh}. We have re-computed these eikonal functions for the
case of two hard partons, that we require for the analysis in this paper, and
found agreement with the results reported in these references.

Using these results for the eikonal functions, we can construct all required
real-emission integrals with constraints provided by the definition of the
zero-jettiness variable. These integrals are complicated, and we make use of the
various internal symmetries to simplify them. For example, we note that upon
relabelling $n \leftrightarrow \bar n$, integrals do not change. This implies
that we have only \emph{two} classes of integrals to consider. The first class
involves integrals where the phase-space constraint is
\begin{equation}
\delta(1 - \sum \limits_{i=1}^{3} \beta_i) \prod \limits_{i=1}^{3}
\theta(\alpha_i - \beta_i).
\end{equation}
We will refer to such integrals as $nnn$ integrals.  The second class of integrals are integrals where the constraint  reads
\begin{equation}
\delta(1 - \sum \limits_{i=1}^{2} \beta_i - \alpha_3) \prod \limits_{i=1}^{2}
\theta(\alpha_i - \beta_i) \; \theta(\beta_3 - \alpha_3).
\end{equation}
We will refer to these integrals as $nn \bar n$ integrals. All integrals that
are needed for the calculation of $S_\tau^{(3)}$ can be written as a linear
combination of $nnn$ and $nn\bar n$ integrals using the freedom to re-name $n
\leftrightarrow \bar n$ and $k_i \leftrightarrow k_j$, $i,j \in \{1,2,3\}$.

To further reduce the number of triple-real integrals that have to be
calculated, we employ integration-by-parts (IBP)
identities~\cite{Tkachov:1981wb,Chetyrkin:1981qh}. To make this technology
applicable to phase-space integrals which contain $\theta$-functions as
constraints, we use its extensions discussed in
Refs.~\cite{Anastasiou:2002yz,Baranowski:2021gxe}. Following
Ref.~\cite{Anastasiou:2002yz} and re-writing phase space integrals as cuts of
loop integrals, we derive the integration-by-parts identities where we encounter
additional terms associated with derivatives of (some) $\theta$-functions. Since
${\rm d} \theta(x)/{\rm d} x = \delta(x)$, we can apply reverse unitarity
\cite{Anastasiou:2002yz} one more time, obtaining a system of equations for
integrals with $\theta$-functions and integrals where one of the
$\theta$-functions is replaced by a $\delta$-function. Selecting the latter, and
writing the integration-by-parts equations for them, we obtain integrals that
contain two $\delta$-functions and one $\theta$-function, in addition to the
original ones. Applying this procedure iteratively, we obtain a list of
integrals that close the system of integration-by-parts equations. These
integrals contain various number of $\theta$- and $\delta$-functions; the
original integrals possess three $\theta$-functions and the "simplest" ones --
three $\delta$-functions.

 Upon generating sufficiently many IBP equations, we obtain a system of linear
equations that contains all integrals that are required to compute the
triple-real emission contribution to the soft function. We then solve this
system and express the required integrals through a set of basis (or master)
integrals. Once master integrals are calculated, the result for the triple-real
emission contribution to the soft function is obtained.

Although the set-up that we describe above is standard, realizing it in practice
turned out to be quite challenging. We list below some of the issues that we
encountered and briefly comment on how we solved them.

\begin{itemize}
  
\item The (usually straightforward) step of \emph{constructing} linear algebraic
  equations for the required triple-real integrals with the help of the
  integration-by-parts method turns out to be non-trivial in this case because
  none of the public computer codes dedicated to processing loop integrals and
  deriving and solving IBP equations for them, can work with $\theta$-function
  constraints.
  
\item A replacement of $\theta$-functions with $\delta$-functions in integrands
  creates difficulties with defining unique integral families since, upon such
  replacements, propagators that define classes of integrals become
  linear-dependent. A multi-variate partial fractioning has to be applied to map
  integrals on the unique set of families. This has to be done on the fly when the
  integration-by-parts identities are constructed.
  
\item Once a large-enough system of linear equations is obtained, we solve it
  using the program
  \texttt{Kira}~\cite{Maierhofer:2017gsa,Maierhofer:2018gpa,Klappert:2020nbg}. To
  obtain the solution, we need to choose preferred master integrals, and we do so
  by using criteria such as the minimal number of $\theta$-functions, the simplest
  structure of propagators etc.
  
\item As explained in Ref.~\cite{Baranowski:2022khd}, not all
  integrals with $\theta$-function constraints are regularized dimensionally.  To ameliorate this problem, we need to introduce
  an analytic regulator which provides
  an additional  challenge for various steps discussed in the previous items. Although the 
  dependence on the analytic regulator cancels in the final expression for the soft function, it significantly complicates
  the reduction to master integrals and their calculation.
  
\item We have attempted to compute the required master integrals analytically.
  In many cases this can be done in a relatively straightforward way provided that
  singular limits of integrands can be identified, subtracted and evaluated
  separately. To calculate finite remainders of divergent master integrals whose
  integrands can be expanded in series in $\ep$, we relied very heavily on the
  program \texttt{HyperInt}~\cite{Panzer:2014caa}. It is not an exaggeration to
  say that without it, obtaining analytic results for the majority of integrals
  would not have been possible.
  
\item Diagrams that describe processes where an off-shell
  gluon splits into three soft  gluons lead to integrals that
  contain the following propagator
  \begin{equation}
    \frac{1}{k_{123}^2} =
    \frac{1}{2 k_1 k_2 + 2 k_1 k_3 + 2k_2 k_3}.
  \end{equation}
  We were unable to compute integrals with such a propagator analytically by
  direct integration. Instead, we modified this propagator by introducing a
  ``mass'' term
  \begin{equation}
    \frac{1}{k_{123}^2} \to
    \frac{1}{k_{123}^2 + m^2},
  \end{equation}
  making  master integrals functions
  of $m^2$.   It is then straightforward to  derive the
  differential equations for $m$-dependent master integrals.\footnote{
    In fact, for this case we have found a way to remove 
    unregulated integrals from the system of 
    integration-by-parts identities which led to a dramatic
    simplification in the IBP reduction required for deriving differential equations.}
  Once the differential equations become available, we solve them numerically,
  starting from the $m^2 \to \infty$ limit, where boundary constants can be
  computed, and propagate these solutions to $m^2 = 0$, where a particular limit
  of mass-dependent integrals, that corresponds to \emph{Taylor series} in $m^2$,
  is needed.
  
\item Calculation of boundary conditions at $m^2 = \infty $ is a highly
  non-trivial problem. Indeed, these \emph{real-emission} integrals possess
  ultraviolet divergencies since the $\delta$-function $\delta(\tau - {\cal T}_0)$
  does not constrain energies of all the partons. This fact results in the
  appearance of ${\cal O}(m^{-4\ep})$ and ${\cal O}(m^{-2\ep})$ ``branches'' in
  $m^2 \to \infty$ integrals, in addition to regular Taylor branches which
  correspond to a ``naive'' expansion of integrands in powers of $1/m^2$.
  
  We have computed all the required
  boundary constants analytically. To reduce the number
  of independent boundary integrals that have to be calculated,
  we simplified the master integrals at $m^2 \to \infty$
  along the lines described in Ref.~\cite{Baranowski:2022vcn},
  and used integration-by-parts identities to derive relations amongst
  \emph{expanded integrals of a specific branch}.
  
\item We have tested the computed master integrals in several ways. First, we
  have constructed a Mellin-Barnes representation for all $m^2 = 0$ integrals,
  required for the calculation of the soft function, and computed them
  numerically. We were able to verify the analytic results, as well as the results
  of $m^2 \to 0$ extrapolation of the solution of the differential equations for a
  large number of highly non-trivial integrals.
  
  We have also checked solutions of the differential equations for master
  integrals at finite $m^2$ by using them to predict results for other integrals
  that can be evaluated by a direct numerical integration. The numerical
  integration is performed using the sector decomposition
  \cite{Binoth:2000ps,Binoth:2003ak} and, by choosing the integrals carefully, it
  is possible to keep the size of the resulting expressions under control. By
  computing many such integrals, we were able to check all the master integrals at
  finite $m^2$.

\item As part of the effort to compute the missing $n n \bar n$ real-emission
  integrals, we have re-calculated the $nnn$ contribution to the soft function at
  N$3$LO, reported earlier in Ref.~\cite{Baranowski:2022vcn} and found complete
  agreement.

\end{itemize}

\section{Renormalization}
\label{sec:renorm}

We are now in position to assemble the final result for the soft function. To
write it in the simplest way possible, we note that $S_\tau$ is a
\emph{distribution} in the variable $\tau$. The leading order term is normalized
to be a $\delta$ function, $\delta(\tau)$, and terms of the type
$\tau^{-1-2n\ep}$ with $n=1,2,3$ etc. appear in higher perturbative orders. In
fact, for a generic $s_{ab}$ and $P$ in the definition of the jettiness
variable, Eq.~(\ref{eq:Tau0def}), the $n$-th order contribution to the soft function
will have a prefactor
\begin{equation}
  \label{eq:XtauDef}
  X_\tau^n = \frac{1}{\tau} \left( \frac{\tau P}{\mu\sqrt{s_{ab}}} \right)^{-2n\ep},
\end{equation}
 where the renormalization scale $\mu$ appears when the bare QCD coupling
 constant is re-written through the renormalized one. We note that the
 ${\overline {\rm MS}}$ renormalization scheme is used throughout this paper. To
 avoid the need to deal with the distributions, it is convenient to work with the
 Laplace transform of the soft function defined as follows
 \begin{equation}
   \label{eq:Slap}
   S =  \int \limits_{0}^{\infty} {\rm d} \tau \; S_\tau \; e^{-\tau u},
 \end{equation}
 where $u$ is the parameter of the Laplace transform. 
 Integrating the quantity in Eq.~(\ref{eq:XtauDef}), we find 
 \begin{equation}
   \label{eq:Xpref}
   \int \limits_{0}^{\infty} {\rm d} \tau \; 
   X_\tau
   e^{-\tau u} 
   = -\frac{e^{-2\ep n \gamma_E} \Gamma(1-2n)}{2n \ep} \; 
   e^{2 n \ep L_S},
 \end{equation}
 where $L_{S} = \log \left(\bar{u} \mu \sqrt{s_{ab}}/ P \right)$,
 $\bar{u} = u e^{\gamma_E}$, and  $\gamma_E$  is the Euler–Mascheroni constant.  
 Hence, we conclude that the Laplace transform  of the soft function
 depends on the parameter $L_S$.

The bare soft function $S$ is a divergent quantity; it requires a dedicated
renormalization. In the Laplace space, this renormalization is multiplicative,
\begin{equation}
  \label{eq:renZs}
  \tilde S = Z_s \; S ,
\end{equation}
where $\tilde S$ is the renormalized soft function that contains no
$1/\ep$-poles, and $Z_s$ is a function of the strong coupling constant
$\alpha_s(\mu)$ and $L_s$.

The renormalization constant $Z_s$ can be determined from the renormalization
group equation \cite{Korchemsky:1993uz,Korchemsky:1992xv}
 \begin{equation}
   \label{eq:ZsRG}
   \mu \frac{{\rm d} }{{\rm d} \mu } \log{Z_s(\mu)} = -4 \Gamma_{\textrm{cusp}} L_S - 2 \gamma^s.
 \end{equation}

The cusp anomalous dimension $\Gamma_{\rm cusp}$ and the soft anomalous
dimension $ \gamma^s$ in eq.~(\ref{eq:ZsRG}) can be found in
Ref.~\cite{Billis:2019vxg}, as extracted from
Refs~\cite{Kramer:1986sg,Korchemsky:1987wg,Matsuura:1987wt,Matsuura:1988sm,Harlander:2000mg,Moch:2004pa,Vogt:2004mw,Gehrmann:2005pd,Moch:2005tm,Moch:2005id}.
We provide them in digital form in an ancillary file provided with this
submission.

The renormalization group equation Eq.~(\ref{eq:ZsRG}) can be solved order by
order in the expansion in the strong coupling constant $\alpha_s$. Once $Z_S$ is
obtained, it is straightforward to use it to predict $1/\ep$-poles of the bare
soft function. The leading pole of the $n$-th order contribution is
$1/\ep^{2n-1}$, which means that for the N$3$LO soft function, there are five
$1/\ep$ poles that the reported calculation should reproduce. It goes without
saying that this should occur independently for different colour and $n_f$
factors, so in practice the number of checks that our calculation has to pass is
quite significant. Indeed, we find that our result for the bare soft function
does reproduce \emph{all} the $1/\ep$ poles that are predicted by the
renormalization group equation.

\section{Results and discussion}
\label{sec:results}

To present the result for the renormalized soft function, 
we write an expansion of its logarithm in powers of $a_s=\alpha_s(\mu)/(4\pi)$
\begin{equation}
  \label{eq:SlogsExp}
  \log \left [ {\tilde S}(L_S) \right ]=
  \sum\limits_{i=1}^\infty \as ^i \sum\limits_{j=0}^{2i} C_{i,j} L_S^j \,.
\end{equation}
At each order in the perturbative expansion, the coefficients of the logarithms
$L_S$ are predicted in terms of cusp and soft anomalous dimensions, and the soft
function at lower perturbative orders. Thus, the soft function can be fully
reconstructed from the expression where the logarithm $L_S$ is set to zero, $L_S
\to 0$. Writing
\begin{equation}
  \label{eq:logS0}
  \log \left [ \tilde   S(0) \right ] = 
  C_R \; {\cal C},
\end{equation}
where $C_R$ is the Casimir operator of the hard emitters ($C_R = C_F \; (C_A)$
for quarks (gluons)), we obtain
\begin{widetext}
\begin{equation}
\begin{split}
  \label{eq:SrenCRfact}
  {\cal C} 
   =
    -\as \pi^2 
  & + \as^2 \left[ \nf \TF \left(\frac{80}{81}
    +\frac{154 \pi^2}{27}
    - \frac{104 \zeta_3}{9} \right)
    - \CA \left( \frac{2140}{80}
    +  \frac{871\pi^2}{54}
    -  \frac{286\zeta_3}{9}
    -  \frac{14\pi^4}{15}\right)\right] \\
    &
      +\as^3 \left[ \nf^2\TF^2 \left(
      \frac{265408}{6561}
      - \frac{400\pi^2}{243}
      - \frac{51904\zeta_3}{243}
      + \frac{328\pi^4}{1215}
      \right)
      + \nf \TF \left(\CF X_{FF}
      +  \CA X_{FA}\right)
      + \CA^2 X_{AA}
      \right],
\end{split}
\end{equation}
\end{widetext}
where the three quantities $X_{FF}, X_{FA}$ and $X_{AA}$, truncated to 16 significant digits, read 
\begin{alignat}{2}
  &X_{FF}=&    68&.9425849800376,
                   \nonumber \\
  &X_{FA}=&   839&.7238523813981, 
                   \label{eq:Xnum} \\
  &X_{AA} =& -753&.7757872704537.
                   \nonumber
\end{alignat}
Eqs.~(\ref{eq:SrenCRfact},\ref{eq:Xnum}) are the main result of the paper. We note
that it should be possible to obtain even more precise numerical results within
our set-up, and eventually reconstruct a fully analytic result using standard
methods as was done in Ref.~\cite{Baranowski:2022khd}. We also note that the
computer-readable result for the renormalized soft function $\tilde{S}(L_S)$
through N3LO can be found in the ancillary files provided with this submission.

 The results reported in Eq.~(\ref{eq:Xnum}) can be compared with earlier attempts
to deduce the non-logarithmic contribution to the soft function using existing
NNLO QCD prediction for thrust in $e^+e^- \to 3~{\rm jets}$
\cite{Gehrmann-DeRidder:2014hxk}, the result for the N3LO hard function for
$e^+e^- \to 2~{\rm jets}$ \cite{Abbate:2010xh} and the result for the N3LO jet
function \cite{Bruser:2018rad}. Using the value of the N3LO non-logarithmic
contribution to the cumulative cross section for thrust in the singular limit
extracted in Ref.~\cite{Monni:2011gb} from a fit to a fixed order prediction
obtained with the \texttt{EERAD3} Monte Carlo program
\cite{Gehrmann-DeRidder:2014hxk}, the ${\cal O}(a_s^3)$ contribution to the soft
function was obtained in Ref.~\cite{Bruser:2018rad}. In our notation, this
result corresponds to the coefficient of the $a_s^3$ term in the expansion of
$\tilde S(0)$ with $C_R = C_F$ and $n_f = 5$. Denoting this coefficient as
$c_{3,qq}$, and using numerical results shown in Eq.~(\ref{eq:Xnum}), we find that
our calculation yields
\begin{equation}
  \label{eq:c3qNF5}
  c_{3,qq}  = -1369.575849.
\end{equation}
The constant $ c_{3,qq}$ in Eq.~(\ref{eq:c3qNF5}) should be contrasted with the value
of $c_{3,qq}$ obtained in Ref.~\cite{Bruser:2018rad} using the result of the
numerical fit reported in Ref.~\cite{Monni:2011gb},
\begin{equation}
  \label{eq:c3qNF5fit}
  c_{3,qq}^{\textrm{fitted}}  = -19988 \pm 1440({\rm stat}) \pm 4000({\rm syst}).
\end{equation}
Since the difference between the two results is substantial, we note that
possible issues with the validity of the fit result obtained with the
\texttt{EERAD3} program were recently pointed out in Ref.~\cite{Bell:2023dqs}.
Interestingly, a significantly smaller value of $c_{3,qq}$ may have important
implications for the determination of the strong coupling constant from the
thrust distribution, pushing it to slightly larger values, see
Ref.~\cite{Bell:2023dqs}.

\section{Conclusion}
\label{sec:concl}

In this letter, we reported the result of the calculation of the N3LO QCD
contribution to the soft function defined using the zero-jettiness variable.
This soft function is the last ingredient that is needed to enable the use of
the zero-jettiness slicing scheme for N3LO computations for colour-singlet
production at the LHC or for N3LO description of the two-jet production in
$e^+e^-$ annihilation. This soft function can also be used to re-sum
zero-jettiness logarithms with N$3$LL accuracy for the thrust distribution in
the vicinity of the two-jet limit, as was recently discussed in
Ref.~\cite{Bell:2023dqs} using a fitted value for $c_{3,qq}$. As pointed out in
that reference, the extraction of $\alpha_s$ from the thrust distribution will
certainly benefit from the exact value of $c_{3,qq}$ obtained in this paper.

Our calculation is the first ever computation of the N3LO soft function for
$N$-jettiness variable, albeit for $N=0$. The next logical step is to attempt to
extend it to the $N=1$ case where, ideally, one would use the result for the
$N=0$ soft function as a boundary condition that contains the most demanding
singular limits. This idea was the gist of the approach employed in
Ref.~\cite{Agarwal:2024gws} for computing the NNLO soft function for arbitrary
$N$, and it would be very interesting to attempt to generalize it to the N$3$LO
case.

\section{Acknowledgments}
During the work of this project, we have benefited from useful conversations
with Arnd Behring and Fabian Lange. 
Parts of the computation reported in this paper were performed at  the HPC system at the Max
Planck Computing and Data Facility (MPCDF) in Garching. 

The research of KM and AP was partially supported by the Deutsche
Forschungsgemeinschaft (DFG, German Research Foundation) under grant
396021762-TRR~257. The research of MD was supported by the European Research
Council (ERC) under the European Union’s research and innovation programme grant
agreement 949279 (ERC Starting Grant HighPHun). The work of DB is supported in
part by the Swiss National Science Foundation (SNSF) under contract
200020$\mathrm{\_}$219367.
\bibliography{sfNNNLO}

\end{document}